# Hard Spheres: Crystallization and Glass Formation

By  P.N. PUSEY [1], E. ZACCARELLI [2], C. VALERIANI [1], E. SANZ [1],
W.C.K. POON [1] and M.E. CATES [1]

[1] *SUPA, School of Physics & Astronomy, The University of Edinburgh, Mayfield Road, Edinburgh, EH9 3JZ, UK*

[2] *Dipartimento di Fisica and CNR-INFM-SOFT, Università di Roma La Sapienza, Piazzale A. Moro 2, 00185 Roma, Italy*

Motivated by old experiments on colloidal suspensions, we report molecular dynamics simulations of assemblies of hard spheres, addressing crystallization and glass formation.  The simulations cover wide ranges of polydispersity $s$ (standard deviation of the particle size distribution divided by its mean) and particle concentration.  No crystallization is observed for $s > 0.07$.  For $0.02 < s < 0.07$, we find that increasing the polydispersity at a given concentration slows down crystal nucleation.  The main effect here is that polydispersity reduces the supersaturation since it tends to stabilise the fluid but to destabilise the crystal.  At a given polydispersity (< 0.07) we find three regimes of nucleation: standard nucleation and growth at concentrations in and slightly above the coexistence region; "spinodal nucleation", where the free energy barrier to nucleation appears to be negligible, at intermediate concentrations; and, at the highest concentrations, a new mechanism, still to be fully understood, which only requires small re-arrangement of the particle positions.  The cross-over between the second and third regimes occurs at a concentration, ~ 58% by volume, where the colloid experiments show a marked change in the nature of the crystals formed and the particle dynamics indicate an "ideal" glass transition.

# 1. Introduction

Assemblies of hard spheres in thermal motion constitute simple models for the behaviour of condensed matter in its various states. The fluid-to-crystal freezing transition of hard spheres was discovered more than 50 years ago in the early computer experiments of Wood and Jacobson (1957) and Alder and Wainwright (1957). Not long afterwards, drawing on their own free-volume theories and the sphere packing experiments of Scott (1960) and Bernal and Mason (1960), Cohen and Turnbull (1959, 1964) suggested that compressing an assembly of hard spheres fast enough to by-pass crystallization should result in a metastable, amorphous, solid "glass" state. In 1986, experiments by Pusey and van Megen (1986) on suspensions of colloidal particles that interact via a steep repulsive potential observed both the freezing transition and, at higher concentrations, glass formation.

Over the last 20 years, further simulations, exploiting the ever increasing power of computers, and further experiments on colloidal systems have advanced our understanding of the details of crystallization and glass formation in this apparently simple system. Nevertheless, significant uncertainties still remain. Here we describe new computer simulations aimed at addressing some of these uncertainties. We focus on two main questions which, to some extent, were already raised by Pusey and van Megen's original experiments: 1. How does the nature of the crystallization process change with increasing concentration? 2. How does polydispersity, a distribution of particle size which is inevitable in the experiments, affect both crystallization and glass formation? A third question, always lurking in the background, is: 3. Do hard spheres *really* show a glass transition? Compared to previous work, our simulations span much wider ranges of concentration and polydispersity.

A faithful simulation of the colloid experiments would take full account of the effects of the solvent both in driving the Brownian motion of the particles and in transmitting hydrodynamic interactions between them. Here we do not attempt that, but rather use simpler molecular dynamics which ignore the solvent and assume Newtonian interactions between the particles. Previous work has suggested that at long times, where the particles/molecules have collided many times, Brownian and Newtonian systems show qualitatively the same phenomenology (e.g. Löwen *et al.* 1991).

# 2. Previous work

## *(a) Colloid experiments*

Figure 1 shows the samples studied by Pusey and van Megen (1986, 1987a). The particles consist of solid cores of amorphous poly(methylmethacrylate) which are coated by thin brushes of a flexible polymer. Compression of these polymer brushes on close approach of two particles results in repulsion described by a steep, nearly hard-sphere, potential. The average radius of the particles is about 320 nm and their polydispersity $s$, defined as the standard deviation of the distribution of radii divided by its mean, is $s \sim 0.05$. The particles are suspended in a mixture of liquids chosen to nearly match their refractive index so that the suspensions are nearly transparent. The samples are illuminated obliquely from behind by a broad beam of white light. Their concentrations, expressed as volume fraction $\phi$ (see below and Table 1), range from about 0.48, sample 2, to nearly 0.64, sample 10.

Figure 1(a) shows the samples immediately after they were mixed by slow tumbling. This process "shear melts" any crystals present and drives the samples into fluid or metastable fluid states. After sitting undisturbed for one day (Fig. 1(b)), sample 2, below the freezing concentration, remains an equilibrium fluid. At higher concentrations, homogeneously-nucleated crystallites are seen as coloured specks due to Bragg reflections. In the coexistence region, samples 3-5, these have settled under gravity to show well-defined interfaces between upper fluid and lower polycrystalline phases. Samples 6 and 7 are filled with small compact crystallites. Sample 8 shows a different kind of crystallization that leads to much larger, irregularly shaped, crystallites. After one day, no crystallization is seen in the two most concentrated samples, 9 and 10.

After four days (Fig. 1(c)), a small amount of crystal has formed through sedimentation at the bottom of the equilibrium fluid sample 2. Sedimentation is also evident at the bottom of samples 3-7 where the compressed crystallites show a different coloured Bragg reflection (more green than red/yellow). Coarsening of the crystallites can also be seen, particularly in sample 7. Sample 8 is now filled with the larger, irregular crystallites, and some crystallization is evident at the top of both samples 9 and 10.

Concentrations by weight of colloidal suspensions are relatively easy to measure. However, for several reasons (see e.g. Segrè *et al.* 1996), determination of volume fractions is more problematic. Pusey and van Megen (1986) used an approach that has been adopted in much of the subsequent experimental work. Extrapolation to zero of the amount of crystal phase in the coexistence region gives the freezing concentration. This concentration (by weight) is then identified with the freezing volume fraction $\phi_F = 0.494$ found in accurate computer simulations of monodisperse (equal-sized) hard spheres (Hoover and Ree 1968). The volume fractions of other samples are then determined by suitable scaling of their weight concentrations. For the samples of Fig. 1, this approach leads to a melting volume fraction of $\phi_M \sim 0.536$, slightly smaller than Hoover and Ree's value, $\phi_M = 0.545$. Sample 7, the last to show small homogeneously-nucleated crystals, has $\phi = 0.577$ and, for sample 8, $\phi = 0.595$.

Subsequent studies, by dynamic light scattering, of the microscopic Brownian dynamics of samples similar to those of Fig. 1(a) – i.e. shear-melted metastable fluids before significant crystallization – showed the appearance of non-decaying plateaux in the dynamic scattering functions as concentration was increased (Pusey and van Megen 1987*b*; van Megen and Underwood 1994). This observation implies the partial freezing-in of density fluctuations and the suppression of long-distance diffusion. The striking coincidence of the concentrations, $\phi \sim 0.58$, at which homogeneously-nucleated crystallization ceases and long-distance diffusion is suppressed led Pusey and van Megen (1987b) to identify the two observations with a glass transition. Thus, although the equilibrium state of samples 9 and 10 in Fig. 1(b) is presumably crystalline, the particles are so tightly packed that they cannot move far enough to form crystal nuclei and the samples are trapped in a glassy state.

*(b)  How does the nature of the crystallization process change with increasing concentration?*

Despite this evidence for a glass transition, it is clear from Fig. 1 that crystallization can take place at concentrations higher than $\phi = 0.58$ (samples 8-10). Perhaps too casually, Pusey and van Megen (1986) suggested that the process at these high concentrations involves heterogeneous nucleation at the meniscus and at cell walls. In fact, in a more careful study of samples around $\phi \sim 0.58$, van Megen and Underwood (1993) found that nucleation also occurs in the bulk of the samples.

Later experiments in the microgravity of space showed that full crystallization, similar to that seen in sample 8 ($\phi = 0.593$), can occur at concentrations as high as 0.62 (Zhu *et al.* 1997). Furthermore, computer simulations also found crystallization at high concentrations (Rintoul and Torquato 1996). This led both groups of authors to suggest that hard spheres do not exhibit a glass transition. Nevertheless, it is clear from the experiments in both micro- and normal gravity that there is a marked change in the nature of the crystallization mechanism at around $\phi = 0.58$. One of the aims of the current simulations is to understand this phenomenon better, and to clarify the connection, if any, with a glass transition.

For completeness, we suggest a possible explanation for the partial crystallization seen in sample 9 (Fig. 1(c)), which has a concentration, $\phi \sim 0.62$, similar to that of the sample which showed full crystallization in microgravity. In sample 9, crystallization started at the top and grew downwards for about 10 days before stopping, leaving the lower half of the sample amorphous. It seems likely that the crystals were nucleated when the concentration at the top was reduced somewhat by sedimentation of the particles. Because particles pack efficiently in the crystal, the crystallization process creates free volume at the crystal-fluid interface, allowing further downward growth. However, because of gravitational settling, a layer of random-close-packed sediment ($\phi \sim 0.64$), in which the particles are essentially immobile, grows from the bottom. When the crystal interface meets this close-packed region, growth ceases.

*(c)  How does polydispersity affect crystallization and glass formation?*

It has been known for a long time that a distribution of size can strongly affect the crystallization behaviour of spherical particles. For example, it is found experimentally that colloidal systems with a polydispersity *s* greater than about 0.10 do not crystallize on an experimentally accessible timescale at any concentration (e.g. Pusey 1987). Likewise, binary mixtures with a size difference of 20% ($s \sim 0.12$), or more, are frequently used to avoid crystallization in simulations of glassy behaviour (e.g. Bernu *et al.* 1987). Earlier simulations which focussed on the effects of polydispersity include: Moriguchi *et al.* (1993, 1995), who considered a system of slightly soft spheres; Auer and Frenkel (2001a, 2001b), who calculated nucleation rates at relatively low concentrations; and Williams *et al.* (2001, 2008), who studied a binary mixture of hard spheres with size ratio 0.9.

Probably the most detailed theoretical study of the equilibrium phase diagram of polydisperse hard spheres is that of Fasolo and Sollich (2004). In Fig. 2, a reproduction of their Fig. 8, the predicted phase boundaries are shown in a volume fraction-polydispersity ($\phi - s$) representation. These results are for a triangular size distribution, though it is likely that they will not differ much for other, relatively narrow, distributions.

We note several features of these predictions. As expected, the monodisperse system, $s = 0$, shows freezing at $\phi_F = 0.494$ and melting at $\phi_M = 0.545$. A polydispersity smaller than about 0.02 has relatively little effect on the predicted behaviour. However, for $s > 0.02$ the freezing and melting lines bend markedly towards higher concentrations, implying that polydispersity stabilizes the fluid but destabilizes the crystal. This observation can be understood as follows. Although counterintuitive at first sight, it is now accepted that equal-sized hard spheres crystallize because, above the melting concentration, they have more free volume for local motions – and thus higher entropy – in the (apparently) ordered crystal rather than in the disordered metastable fluid (e.g. Frenkel 1993; Ackerson 1993). In other words, particles pack more efficiently in the crystal. When polydispersity is introduced, the fluid is stabilized because more efficient packing is achieved by distributing the different sized particles among the different sized spaces in the random structure. However, in the crystal, particles of all sizes occupy cells of one size, set by the crystal lattice, resulting in a less efficient packing (Phan *et al.* 1998).

Figure 2 predicts that a single crystalline phase, incorporating particles of all sizes, is stable only for polydispersities smaller than about 0.065. At higher polydispersities and high concentrations, multiple crystal phases are predicted, formed from narrower sub-populations of the parent population, each having a different mean size.

The particles used for Fig. 1 had a polydispersity of about 0.05, for which the freezing concentration is predicted to be $\phi_F = 0.508$ (Fig. 2). As explained in Sec. 2(a), in that early work (Pusey and van Megen 1986), the freezing concentration was assumed to take the value, 0.494, for a monodisperse system and other sample concentrations were scaled in terms of this value. Now we see that the corrected or "true" concentrations, $\phi_{cor}$, of these samples can be obtained by multiplying the nominal values, given above and in Pusey & van Megen (1986), by the factor $0.508/0.494 = 1.028$ (Table 1).

Table 1: Concentrations of samples in Fig. 1: $\phi$, nominal concentration; $\phi_{cor}$, corrected concentration; $\phi_{cor} - \phi_F$, supersaturation.

| Sample | 2 | 3 | 4 | 5 | 6 | 7 | 8 | 9 | 10 |
|---|---|---|---|---|---|---|---|---|---|
| $\phi$ | 0.478 | 0.503 | 0.511 | 0.527 | 0.553 | 0.577 | 0.595 | 0.620 | 0.637 |
| $\phi_{cor}$ | 0.491 | 0.517 | 0.525 | 0.542 | 0.568 | 0.593 | 0.611 | 0.637 | 0.654 |
| $\phi_{cor} - \phi_F$ | | 0.009 | 0.017 | 0.034 | 0.060 | 0.085 | 0.103 | 0.129 | 0.146 |

At $s = 0.05$, a single solid phase is predicted in Fig. 2 for concentrations between 0.552 and 0.609. Interestingly, the corrected concentration of sample 8, the highest concentration sample to show full crystallization (Fig. 1), is close to 0.609.

## 3. Simulation Details

We report molecular dynamics simulations of a system of $N = 2000$ particles in the *NVT* ensemble. An event-driven algorithm has been implemented for particles interacting via hard potentials. Between collisions, particles move along straight lines with constant velocities. When the particles touch, i.e. when the distance between the particle centres becomes equal to the sum of their radii, the velocities of the interacting particles change instantaneously according to classical laws of elastic collision. The algorithm calculates the time to the next collision in the system and propagates the trajectory from one collision to the next one (e.g. Rapaport 1995; Zaccarelli et al. 2002). In order to simulate a bulk system, we accommodate the particles in a cubic box, and periodically repeat the box in all three directions (e.g. Frenkel and Smit 2002; Allen and Tildesley 1988).

Time is measured in units of $\overline{\sigma}(m/k_B T)^{1/2}$, where $\overline{\sigma}$ is the average diameter of the particles and $m$ their mass; $k_B$ is the Boltzmann constant and $T$ the temperature. (Thus in one reduced time unit, a free particle would move a distance of about one diameter.) When dealing with polydisperse systems, particle diameters are chosen according to a 31-component discrete Gaussian distribution with relative standard deviation $s$. The packing fraction is defined as $\phi \equiv \pi N \overline{\sigma}^3 / 6V$ (*V* being the system volume) in order to avoid ambiguities when comparing state points with different size distributions. We have studied extended ranges of polydispersities, from monodisperse samples, $s = 0$, to $s = 0.085$, and packing fractions, from $\phi = 0.54$ to $\phi = 0.63$. A state point is defined by given values of $s$ and $\phi$. All the simulations are run for a maximum (reduced) time $t_{MAX} = 10^5$. To improve the statistics of the results, we have considered 5 different runs for each state point, every run starting from an independent initial configuration whose fraction of crystalline particles is checked to be lower than 5%. In order to create an initial configuration for every packing fraction, we generate a random distribution of points in a cubic box and let their diameters slowly grow, according to the chosen polydispersity, in small time steps to avoid overlaps. In this way we prepare configurations at the highest packing fraction ($\phi \approx 0.64$). To generate configurations with $\phi$ smaller than 0.64, we allow the latter configurations to equilibrate in a larger box.

We then monitor properties like the mean-square displacement, the pressure and the fraction of crystalline particles. The mean-square displacement is calculated from

$$\left\langle \Delta r^2(t) \right\rangle = \frac{1}{N} \sum_{j=1}^{N} \left[ \boldsymbol{r}_i(t) - \boldsymbol{r}_i(0) \right]^2,$$

where $\boldsymbol{r}_i(t)$ is the position of particle $i$, in units of the average particle diameter, at reduced time $t$. The (reduced) pressure $p$ is quoted in units of $k_B T / \overline{\sigma}^3$. (Note that, for a monodisperse system, the pressure in these units is equal to $\left(6\phi/\pi\right)Z$ where $Z$ is the "compressibility factor", the pressure in ideal gas units, $Nk_BT/V$ .)

The fraction of crystalline particles is defined according to a rotationally invariant local bond order parameter $d_6$ (Steinhardt *et al.* 1983; van Duijneveldt and Frenkel 1992; ten Wolde *et al.* 1996), known to be a useful tool when the growing crystal has a closed packed structure. The idea is to calculate for each particle a complex vector $q_6(i)$, whose components $m$ depend on the relative orientation of particle $i$ with respect to its neighbouring particles. Each of the 13 components of the vector associated with the $i^{th}$ particle is given by:

$$q_{6,m}(i) = \frac{\frac{1}{N_b(i)}\left(\sum_{j=1}^{N_b(i)}\Psi_{6,m}\left(\theta_{ij},\phi_{ij}\right)\right)}{\left(\sum_{m=-6}^{6}q_{6,m}(i)\cdot q^*{}_{6,m}(i)\right)^{1/2}}, \qquad m=\left[-6,6\right]$$

where $N_b(i)$ is the number of neighbours of the particle $i$ (all particles $j$ within a cut-off distance of $1.4\,\overline{\sigma}$ ), and $\Psi_{6,m}\left(\theta,\phi\right)$ are the spherical harmonics of order 6. Then we compute the rotationally invariant bond order parameter $d_6$ by calculating the scalar product between each particle and its neighbours:

$$d_6(i,j) = \sum_{m=-6}^{6} q_{6,m}(i)\cdot q^*{}_{6,m}\left(j\right).$$

$d_6(i,j)$ is a normalised quantity correlating the local environments of neighbouring particles, it is a real number and is defined in the range $-1\le d_6(i,j)\le 1$ . For example, in a perfect face-centred-cubic crystal, all the particles have the same environment and, therefore, the dot product between the vectors associated with any pair of particles is 1. The dot product decreases when thermal vibrations are present but, on average, it is close to one if particles have a solid-like environment, and around zero if particles have a liquid-like environment. We consider that particles i and j have a "solid connection" if their $d_6(i,j)$ exceeds 0.7. A particle is labelled as solid-like if it has at least 6 solid connections. Finally we define the degree of crystallinity $X(t)$, or simply the "crystallinity", of a sample as the number of particles in solid-like environments at a given time divided by the total number $N$.

To assess whether a state point is solid-like, we monitor its pressure and degree of crystallinity over time: if within the simulated $t_{MAX}$ at least one of the independent runs displays crystallization, the chosen state point corresponds to a solid-like structure. When crystallization does not happen within the chosen $t_{MAX}$, the system remains amorphous: we can thus monitor its dynamics and determine whether it approaches a glassy state.

## 4. Results

### (a) Crystalline or amorphous?

For the problem under consideration, probably the simplest question is whether or not a system crystallizes during an experiment of a certain duration. Figure 3 shows the results of such an analysis, where the crosses indicate crystallization and the circles indicate systems that remained amorphous. As mentioned above, for each volume fraction and polydispersity, five runs were studied for durations (in reduced time) of $t_{MAX} = 10^5$. The criterion for crystallization, leading to a cross in Fig. 3, was that at least one run should show significant, though not necessarily complete, crystallization.

At relatively low volume fractions, $\phi < 0.54$, in the fluid-crystal coexistence region, no crystallization is observed, presumably because the typical time for crystal nucleation is longer than the run time. For the monodisperse system, $s = 0$, crystallization is observed from $\phi = 0.54$ all the way up to $\phi = 0.62$. At larger polydispersities the range of volume fraction over which crystallization is observed decreases, again implying increasingly longer nucleation times. It is interesting that the most polydisperse system to crystallize, $s = 0.07$ and $\phi = 0.58$, lies close to the upper limit of the one-phase crystal region of Fig. 2.

### (b) Pressure, crystallinity and mean-square displacement

We also performed a more detailed analysis of these simulations, following the pressure, the degree of crystallinity, and the mean-square displacement of the particles as functions of time (see Sec. 3). Figure 4 shows typical results: a monodisperse system, $s = 0$ at three volume fractions, 0.54, 0.58 and 0.61, which span the apparent glass transition observed in the experiments of Fig. 1 (Fig. 4(a-c)); and analogous results for a polydispersity of 0.05, roughly that of the colloid experiments, (Fig. 4(d-f)). The occurrence of crystallization is indicated by a relatively fast increase in the degree of crystallinity and an accompanying drop in the pressure. We define a nucleation time $\tau$, indicated by arrows in Fig. 4, as the time when the crystallinity reaches 0.2 (Figs. 4(b) and (e)).

The (reduced) pressure for the monodisperse system at $\phi = 0.54$, Fig. 4(a), remains virtually constant for a long time before dropping abruptly when the system crystallizes. The initial value of the pressure is close to that, $p = 17.7$, predicted by the Carnahan-Starling (1969) expression, known to be accurate for hard spheres in their equilibrium fluid state, and the final pressure is close to that, $p = 11.8$, predicted for fluid-crystal coexistence.

For this same system, the degree of crystallinity, Fig. 4(b), fluctuates between 0 and about 0.03, reflecting the growth and dissolution of sub-critical nuclei. At $t \sim 5000$, the crystallinity suddenly jumps to about 0.90. Presumably a nucleus larger than the critical size has formed and has grown rapidly until the equilibrium coexistence condition of $\sim 90\%$ crystal, expected for $\phi = 0.54$, is reached. The maximum crystallinity before growth of 0.03 suggests that the critical nucleus must contain somewhat more than $0.03N = 60$ particles ($N = 2000$). This value compares favourably with that, $\sim 70$ particles, found in the earlier simulations of Auer and Frenkel (2001a) at $\phi = 0.534$.

The mean-square displacement for this system is shown in Fig. 4(c). Short-time ballistic motion, $\left\langle \Delta r^2(t) \right\rangle \propto t^2$, crosses over into diffusive behaviour, $\left\langle \Delta r^2(t) \right\rangle \propto t$, at longer times. At the nucleation time $\tau \sim 5000$, $\left\langle \Delta r^2(\tau) \right\rangle \approx 100$, implying a root-mean-square (rms) displacement $\left\langle \Delta r^2(\tau) \right\rangle^{1/2} \approx 10$ so that a typical particle has diffused a distance of about 10 particle diameters.

These results for $s = 0$ and $\phi = 0.54$ are consistent with a conventional nucleation-and-growth picture. The system remains in a well-defined, essentially stationary, metastable fluid state until, at random, a nucleus exceeding the critical size is formed, leading to rapid crystallization. As we will now see, at higher concentrations it is a different story.

At $\phi = 0.58$, the pressure decreases somewhat before crystallization occurs at $\tau \sim 200$, leading to about 98% crystallinity. The pressure of the amorphous state stays well above the Carnahan-Starling value of $p = 25.7$. The pressure of the crystal is also significantly higher than the expected equilibrium value, $p = 15.2$, given by the Hall (1972) expression, perhaps because of defects (Phan et al. 1998). The crystallinity at $\phi = 0.58$ grows more or less continuously, slowly at first, then more rapidly after $t \sim 100$. Unlike the behaviour of the more dilute system, there is little evidence of sub-critical nuclei dissolving. The rms displacement for $\phi = 0.58$ reaches $\left\langle \Delta r^2(\tau) \right\rangle^{1/2} \approx 0.6$ at the nucleation time $\tau \sim 200$.

The system at $\phi = 0.61$ shows broadly similar behaviour to that at $\phi = 0.58$, except that the contrast with the more dilute system, $\phi = 0.54$, is more marked. The pressure of the amorphous state is much larger than the Carnahan-Starling value, $p = 34.5$, and it decreases, or ages, significantly with time. The crystallinity shows an initial slow increase followed by more rapid growth after $t \sim 5000$ (though crystallization is not complete in the duration of the run). At the nucleation time $\tau = 5000$, the rms displacement is about 0.4.

The main difference between the results for the monodisperse system and those for a poydispersity $s = 0.05$ (Fig. 4(d-f)) is that, in the latter case, no crystallization is observed for $\phi = 0.54$ and 0.61, and that the crystallization for $\phi = 0.58$ is delayed in time by a factor of more than 10. It is interesting that, for $\phi = 0.61$, aging of the pressure seems to cease around $t = 10^4$ and that the crystallinity actually *decreases* a little with time. At $\phi = 0.54$, the growth and decay of sub-critical nuclei is again evident though their maximum size is much smaller than for the monodisperse system. Note also that, for the reasons given in Sec. 2(c), the pressures of the polydisperse metastable fluids are slightly smaller than those of the corresponding monodisperse systems (compare Figs. 4(a) and 4(d)), whereas the pressure of the polydisperse crystal at $\phi = 0.58$ is slightly larger.

*(c) An ideal glass transition?*

A striking feature of Figs. 4(c) and 4(f) is that, at each concentration, the mean-square displacements, measured before crystallization, are almost the same for the monodisperse system and the $s = 0.05$ polydisperse system; this, of course, implies that the particle dynamics are hardly affected by polydispersity. Elsewhere we have undertaken more detailed analysis of the mean-square displacements of all the systems in Fig. 3 which did not crystallize (this includes some runs marked X in Fig. 3) (Zaccarelli *et al.* 2009). Up to $\phi = 0.59$, it was possible to reach the long-time region, $\left\langle \Delta r^2(t) \right\rangle \propto Dt$, and so derive long-time diffusion coefficients $D$. This analysis confirmed that, at a given concentration, $D$ is virtually independent of polydispersity. Furthermore, we found that the concentration dependence of $D$ could be fitted to the expression $D \propto (\phi - \phi_G)^\gamma$, giving $\gamma \sim 2.15$ and $\phi_G \approx 0.585$. This power-law form of concentration dependence is predicted in the "ideal" glass transition scenario of mode-coupling theory in which activated hopping is neglected (e.g. Götze and Sjögren 1992). It is striking that this value of $\phi_G$, the ideal glass transition concentration, is close to that where the change in the crystallization mechanism and the emergence of non-ergodicity are seen in the colloid experiments (Sec. 2(a)).

*(d) Nucleation times*

Figure 5(a) shows crystal nucleation times as functions of volume fraction and polydispersity. The results are averages of five runs of duration $t = 10^5$ for each system, and, as noted in Sec. 4(b), the nucleation time $\tau$ is taken to be the time when crystallinity reaches 0.2. Qualitatively, the results are consistent with colloid experiments (e.g. Harland and van Megen 1997). At low concentrations, nucleation is slow because of small supersaturation (a small thermodynamic driving force); at high concentrations it slows again because of the slowing particle dynamics (see above); the fastest nucleation occurs around $\phi = 0.56$. Polydispersities less than about 0.02 do not much affect the nucleation rate, but for larger polydispersities there is a strong slowing down, e.g. a factor of 10 or more for $s = 0.05$.

As pointed out in Sec. 2(c), for $s > 0.02$ the freezing line bends strongly towards higher concentrations (Fig. 2). Thus, at a given volume fraction one would expect the thermodynamic driving force for nucleation to decrease as the polydispersity increases. The simulations of Auer and Frenkel (2001b) show that the difference between the chemical potentials of the metastable fluid and the crystal (that determines the driving force) is almost linear in $\phi - \phi_F$ and nearly independent of polydispersity, at least up to $s = 0.05$. Thus, by plotting nucleation time against $\phi - \phi_F$, Fig. 5(b), we compare data at different polydispersities but essentially the same supersaturation. This operation provides some collapse of the data at low concentration, but still leaves an apparently significant effect of polydispersity.

It is, however, now necessary to recognise that, in Fig 5(b), data shown with the same $\phi - \phi_F$ but different polydispersity also have different dynamics (different diffusion rates) since, as discussed in the previous section, the dynamics are strongly dependent on actual concentration $\phi$, but are almost independent of polydispersity at a given $\phi$. Thus, in Fig. 5(c), we replot the data of Fig. 5(b), scaling the nucleation time by the time $\tau_d$ that a particle takes to diffuse one diameter (i.e $\left\langle \Delta r^2(\tau_d) \right\rangle^{1/2} = 1$). Figure 5(c) is therefore a representation of the data of Fig. 5(a) in which the dependence of supersaturation on $s$ and the dependence of dynamics on $\phi$ are both effectively scaled out. We now see a near complete superposition of the results for $s$ = 0 to 0.03, though a relatively small effect of polydispersity remains for $s$ = 0.04 and 0.05.

Before discussing these results, we digress briefly on two relevant issues.

In order to obtain reliable estimates of nucleation times, one clearly requires the system to be large enough that several nucleation events occur during the simulation, leading to several crystallites in the final, fully crystalline, state. To investigate possible dependence on system size, we performed a few simulations of monodisperse systems with a much larger number of particles, either $N$ = 100,000 or $N$ = 54,000. For $\phi$ = 0.54, we indeed found that the nucleation time was about 10 times smaller in the large systems as compared to the small one with $N$ = 2000. This finding is consistent with the observations, at this concentration, of quite large crystallites in the experiment, e.g. sample 5 of Fig. 1(b), and of the growth and decay of a single sub-critical nucleus, discussed in Sec. 3(b), in the simulations with $N$ = 2000. However, at $\phi$ = 0.56 and $\phi$ = 0.58, we found no significant difference between nucleation times measured in large and small systems, implying that several nucleation events occur even in the small system, consistent with the much smaller crystallites observed experimentally, e.g. sample 7 of Fig. 1(b). Interestingly, at $\phi$ = 0.61, where the experiments again show large crystallites, sample 8 of Fig. 1(b), there was not a big difference between nucleation times measured in the small and large systems. Note that reducing the nucleation times by a factor of 10 at the lowest supersaturations in Fig. 5(c), to allow for the system size dependence observed at $\phi$ = 0.54, does not affect the conclusions to be drawn in Sec. 5(a).

The second issue concerns a misleading impression that has been left in the literature for several years. Auer and Frenkel (2001a) calculated nucleation rates for hard spheres, both monodisperse and with a polydispersity of 0.05. As we have found, the polydisperse system showed slower nucleation at a given concentration. Auer and Frenkel pointed out that, as noted above, the main cause was a smaller supersaturation for the polydisperse system, leading to a larger free energy barrier for nucleation. However, when they came to compare their predictions with several sets of experimental data on colloidal systems with polydispersity around 0.05, they understandably used the values of volume fraction quoted in the experimental reports. What they (again understandably) failed to realise was that these experimental volume fractions were calculated assuming freezing to occur at $\phi_F$ = 0.494, the value for a monodisperse system (see Sec. 2(a)). In fact, as explained in Sec. 2(c), the appropriate value for $s$ = 0.05 is $\phi_F$ = 0.508. Thus, in Fig. 2 of Auer and Frenkel (2001a), which compares experimental and theoretical nucleation rates, all the experimental data should be shifted to higher volume fractions by the factor 0.508/0.494 = 1.028. This operation leads to quite reasonable agreement between experiment and theory for $s$ = 0.05, rather than the large discrepancies (by several orders of magnitude) that Auer and Frenkel appeared to find.

## 5. Discussion

### (a) Nucleation mechanisms

Figure 5(c) is probably the most striking of the results that we have presented here. It implies that, when account is taken the effects of supersaturation and particle dynamics, polydispersity, at least up to $s \sim 0.05$, appears to have little influence on the mechanisms of crystal nucleation. Furthermore, the nucleation behaviour appears to fall into three regimes: at small supersaturation $\phi - \phi_F$, particles diffuse several diameters before nucleation takes place, $\tau/\tau_d > 1$; at intermediate $\phi - \phi_F$ only a motion of about one diameter is required, $\tau/\tau_d \approx 1$; and at large $\phi - \phi_F$ crystallization requires only small local displacements of the particles, $\tau/\tau_d << 1$.

The first regime, which extends up to about $\phi - \phi_F = 0.055$ (or $\phi \sim 0.55$ for a monodisperse system), can be identified as conventional nucleation and growth. Here, as described in Sec. 4(b), a nucleus must, at random, grow large enough to overcome a free-energy barrier, determined by a competition between bulk and interface free energies, before macroscopic crystallization takes place. Auer and Frenkel (2001a) have calculated the magnitudes of these barriers for concentrations between 0.521 and 0.534 (monodisperse system). Over this small increase of concentration they find that the height of the barrier drops rapidly. An approximate extrapolation of their results suggests that the barrier height could become small compared to $k_B T$ at a concentration near $\phi = 0.55$-0.56.

At around this concentration, at which $\phi - \phi_F \sim 0.06$, we enter the second regime of Fig. 5(c), where nucleation only requires particles to move about one diameter. We suggest that this corresponds to "spinodal nucleation", already discussed by several authors (e.g. Klein and Leyvraz 1986; Cavagna *et al.* 2005; Trudu *et al.* 2006). The characteristics of this regime are that the driving force for crystallization is large and that there is essentially no free-energy barrier to be overcome. Thus nucleation can start immediately and local re-arrangement of particle positions is all that is necessary to form a crystal. In agreement with this interpretation, some authors (e.g. Trudu *et al.* 2006) have called the point at which the relaxation time of the metastable liquid (which we can take as $\tau_d$) starts to exceed the nucleation time, i.e. $\tau/\tau_d \leq 1$, the "kinetic spinodal".

This spinodal regime in Fig. 5(c) extends to $\phi - \phi_F \sim 0.075$, after which the nucleation time $\tau$ becomes increasingly smaller than the relaxation time $\tau_d$, implying that, at these high concentrations, crystallization can still take place even though the particles, on average, move only a fraction of one diameter. This regime of crystallization does not appear to have been discussed before. It will take further detailed analysis of our simulations to understand properly the mechanisms involved.

*(b) How does the nature of the crystallization process change with increasing concentration?*

Our simulations suggest that crystallization in hard-sphere systems proceeds via three distinct mechanisms. It is interesting to enquire whether there is a connection between these and the nature of the crystallization observed in the experiments, Fig. 1. The standard nucleation-and-growth regime extends to $\phi - \phi_F \approx 0.055$ (above), between samples 5 and 6 in Fig. 1 (see Table 1 for values of $\phi - \phi_F$). Figure 1 shows no obvious change at this concentration. However, moving into the spinodal nucleation regime, samples 6 and 7, we see that the crystallite size decreases markedly with increasing concentration (Fig. 1(b)). This feature, demonstrated more clearly by the detailed experiments of van Megen and Underwood (1993), was highlighted by Schätzel and Ackerson (1993) who pointed out that the crystallite size actually seems to extrapolate to zero at $\phi - \phi_F \approx 0.086$. Small crystallites in the fully crystalline state of course imply a high spatial density of nuclei earlier in the crystallization process. This is at least consistent with the idea of spinodal nucleation in that, in the absence of a free-energy barrier, increasing supersaturation should lead to nucleation on an ever decreasing spatial scale.

The cross-over to large irregular crystals occurs between samples 7 and 8 at $\phi - \phi_F \approx 0.085$. This corresponds to a slightly higher concentration than that, $\phi - \phi_F \approx 0.075$, of the onset of the third regime seen in Fig. 5(c). Nevertheless it is tempting to associate the dramatic change in the crystallization mechanisms between samples 7 and 8 with a change from spinodal nucleation to the new mechanism described in Sec. 5(a).

*(c) How does polydispersity affect crystallization and glass formation?*

The results of Fig. 3, showing no crystallization above a polydispersity of 0.07, are consistent with the existence of a "terminal" polydispersity above which crystals cannot exist without fractionating into multiple species (e.g. Barrat and Hansen 1986; Pusey 1987; Bartlett and Warren 1999; Fasolo and Sollich 2004). Of course, longer simulation runs could show crystallization at states marked amorphous in Fig. 3. In this context we note that, at polydispersity 0.05, the simulations show no crystallization for $\phi > 0.59$, whereas, in the colloid experiments, the crystalline sample 8 has a corrected volume fraction $\phi_{cor} = 0.611$.

At first sight, the plot of nucleation times in Fig. 5(a) seems to imply a large effect of polydispersity for s < 0.07. However, as explained in Sec. 4(d), when account is taken of the dependence of supersaturation on polydispersity and the dependence of particle diffusion rates on actual concentration, these data nearly collapse onto a universal curve (Fig. 5(c)). This suggests that the basic mechanisms of crystallization are hardly affected by polydispersities less than about 0.06.

We note, however, that, after this scaling process, there does appear to be a small residual effect of polydispersity in Fig. 5(c): the data for $s$ = 0.04 and 0.05 lie slightly above the "master" curve, implying slower nucleation. One possible explanation for this is that crystallization at larger polydispersities may involve some fractionation. For example, the initial nucleation might involve just a sub-population of the particles. Any such selection of species needs particles to move over distances comparable to the size of the crystal nucleus, a slow process. A number of experiments by van Megen and co-workers on colloidal systems with differing size distributions suggest that fractionation can indeed be important (Henderson *et al.* 1996; Martin *et al.* 2003; Schöpe *et al.* 2006, 2007). Williams *et al.* (2001, 2008) have observed fractionation in simulations of a binary hard-sphere mixture. In related theoretical work, Evans and Holmes (2001) have pointed out that fractionation can also occur for kinetic reasons: the initial nucleus could prefer smaller-than-average particles simply because they diffuse faster. Furthermore, once a non-equilibrium crystal structure is formed, it will persist because of very low diffusion rates in the crystal.

### *(d) Do hard spheres really show a glass transition?*

As a material is compressed (or cooled), two indications of a glass transition are rapidly slowing dynamics, leading to non-ergodicity, and the onset of aging. For hard spheres, our simulations verify and extend what was already found in the colloid experiments. At all polydispersities that we have studied, particle dynamics slow dramatically at a concentration around $\phi$ = 0.58 - 0.59, consistent with an "ideal" glass transition in the mode-coupling sense. Aging of the systems at $\phi > 0.58$ is evident in the slow decrease of the pressure with time in the plots of Figs. 4 (a) and (d); elsewhere we will also describe the slowing of particle dynamics with increasing age at $\phi > 0.58$ (Zaccarelli *et al.* 2009).

Thus, in both simulations and colloid experiments, hard spheres do show glass-like behaviour at concentrations well below that, $\phi \sim 0.64$, of random close packing. However, it is also clear, again from both simulation and experiment, that, at least for polydispersities less than about 0.07, the systems can crystallize at concentrations considerably larger than that of the apparent glass transition. An important finding of this paper is that there appears to be a new mechanism, beyond classical nucleation and growth and spinodal nucleation, which allows this crystallization. (In more complex systems, such as those composed of non-spherical or network-forming molecules, it seems likely that this new route to crystallization is suppressed, allowing long-lived glasses.) Thus a possibility is that hard spheres *do* form glasses, but that the glasses can crystallize relatively easily through a new mechanism still to be fully understood.

Future work will analyse these simulations in more detail and will provide a more complete comparison with the existing experimental, theoretical and simulation literature.

### Acknowledgements

We thank Giuseppe Foffi for his contributions to the early stages of this work. The simulations made use of the resources provided by the Edinburgh Compute and Data Facility (ECDF) (http://www.ecdf.ed.ac.uk). The ECDF is partially supported by the eDIKT initiative (http://www.edikt.org.uk). We acknowledge grants from the UK Engineering and Physical Sciences Research Council (Edinburgh Soft Matter and Statistical Physics Programme Grant, EP/E030173) and the European Union (Arrested Matter Marie Curie Research Training Network, MRTN-CT-2003-504712, and SoftComp Network of Excellence, NMP3-CT-2004-502235). W.C.K.P. holds an EPSRC Senior Fellowship (EP/D071070). M.E.C. holds a Royal Society Research Professorship.

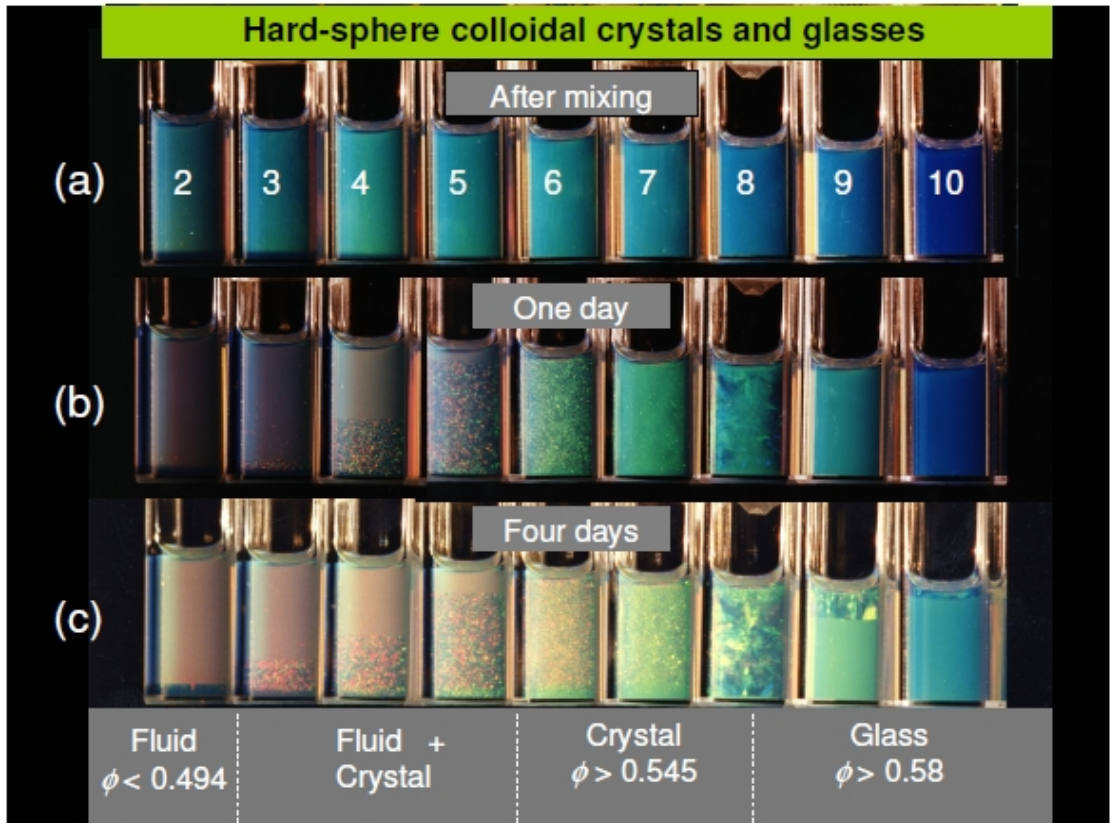

Figure 1

Phase behaviour of hard-sphere colloids. This compilation is formatted more coherently than the limited selection of Pusey and van Megen (1986) and the black and white pictures of Pusey and van Megen (1987*a*); to avoid confusion we retain the original numbering of the samples. (a) Immediately after mixing; sample 2 is equilibrium fluid, others are in metastable fluid/glassy states. (b) After one day. (c) After four days. See text for further description.

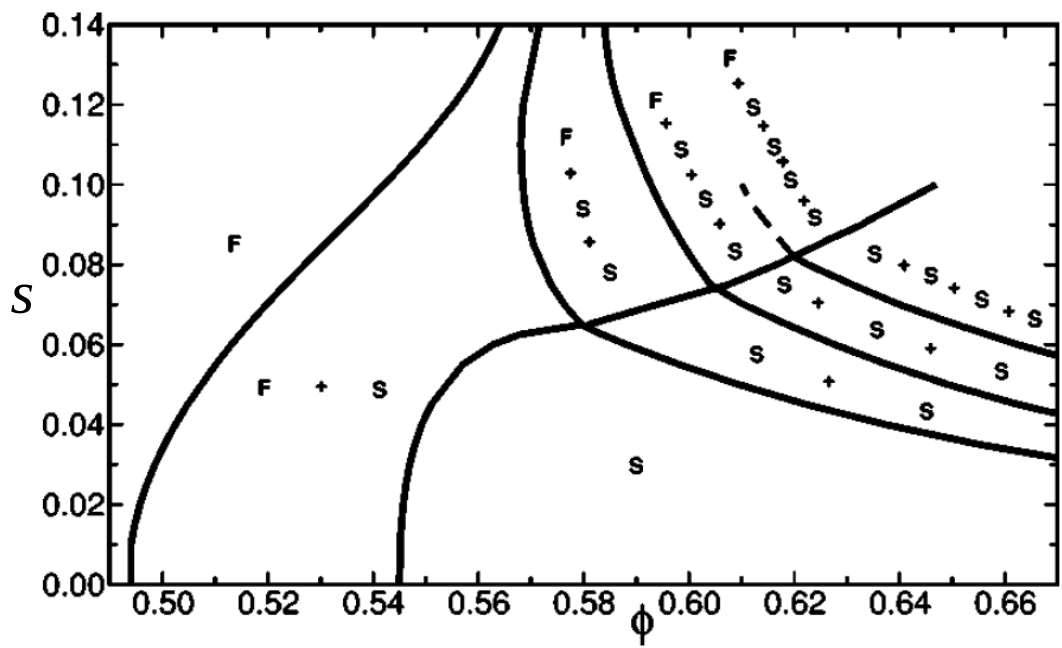

Figure 2

Theoretical phase diagram, in volume fraction-polydispersity, $\phi$ - $s$, representation, for polydisperse hard spheres with a triangular size distribution. F indicates fluid; S indicates solid (crystal); + indicates phase coexistence. Note the prediction of multiple solid phases, with different sub-populations of particles, at large concentrations and polydispersities. (Reprinted with permission from Fasolo and Sollich (2004)).

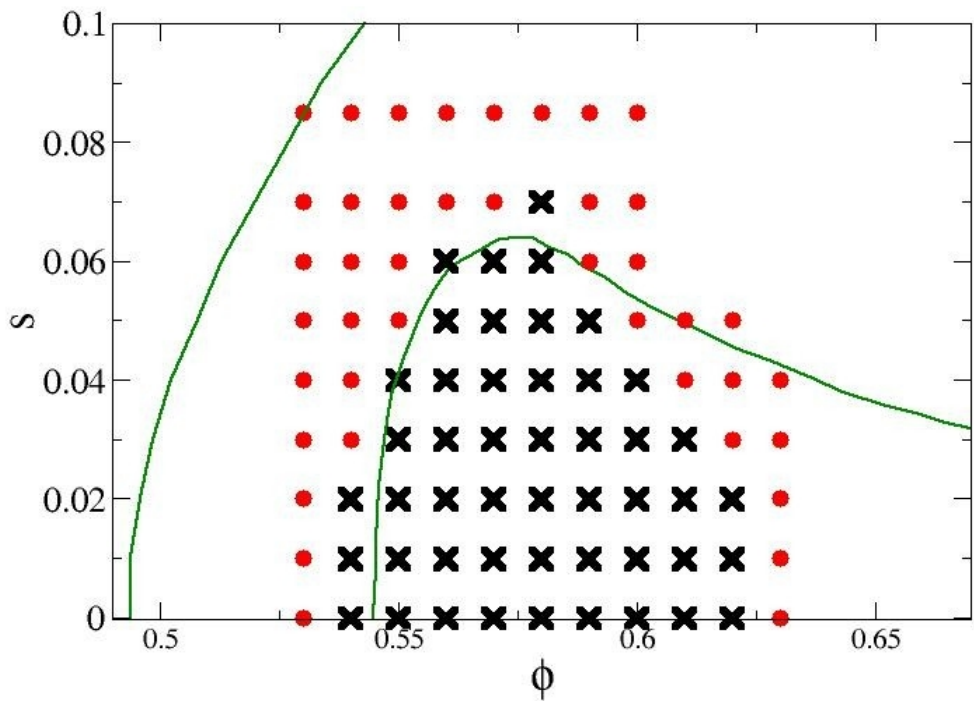

Figure 3

Crystallization diagram from simulations in volume fraction-polydispersity, $\phi$ - $s$, representation. A cross implies that at least one of five independent runs of duration $10^5$ reduced time units at the appropriate state point exhibited crystallization; filled circles indicate no crystallization. The left-hand solid line is the freezing line taken from Fig. 2; the right-hand solid line up to $\phi \sim 0.58$ is the melting line from Fig. 2, above $\phi = 0.58$ it is the line separating S and S + S states.

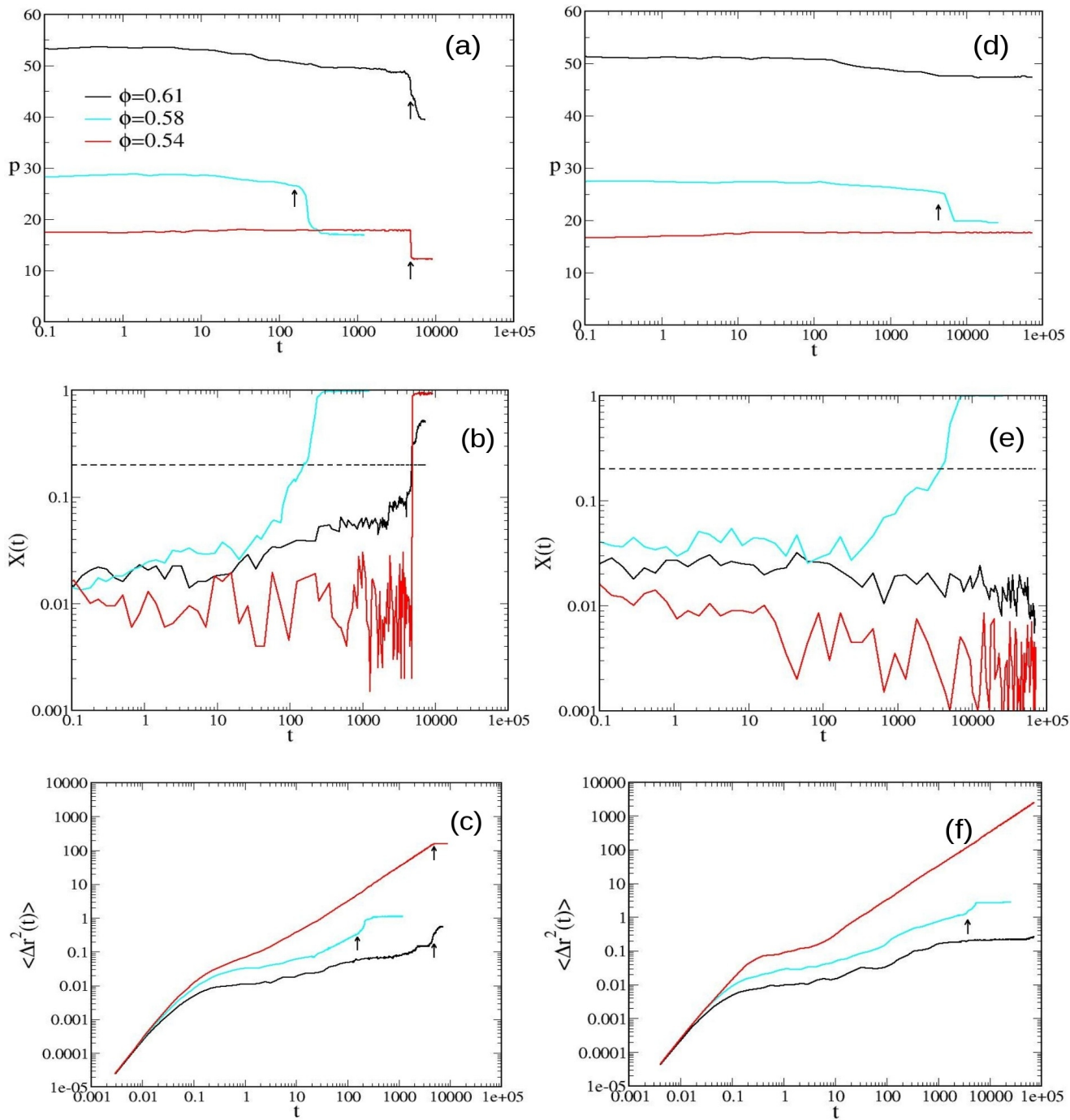

Figure 4

Analysis of the numerical simulations for the monodisperse (a-c) and for the 0.05 polydispersity system (d-f) at three different packing fractions $\phi =$ 0.54, 0.58 and 0.61 as indicated. (a) and (d): reduced pressure $p$ as a function of time for the monodisperse and the polydisperse case, respectively. (b) and (e): degree of crystallinity $X(t)$ as a function of time. (c) and (f): mean-square displacement as a function of time. The arrows in (a), (c), (d) and (f) indicate for each packing fraction $\phi$ the nucleation time $\tau$, when the crystallinity $X(t)$ reaches 0.20.

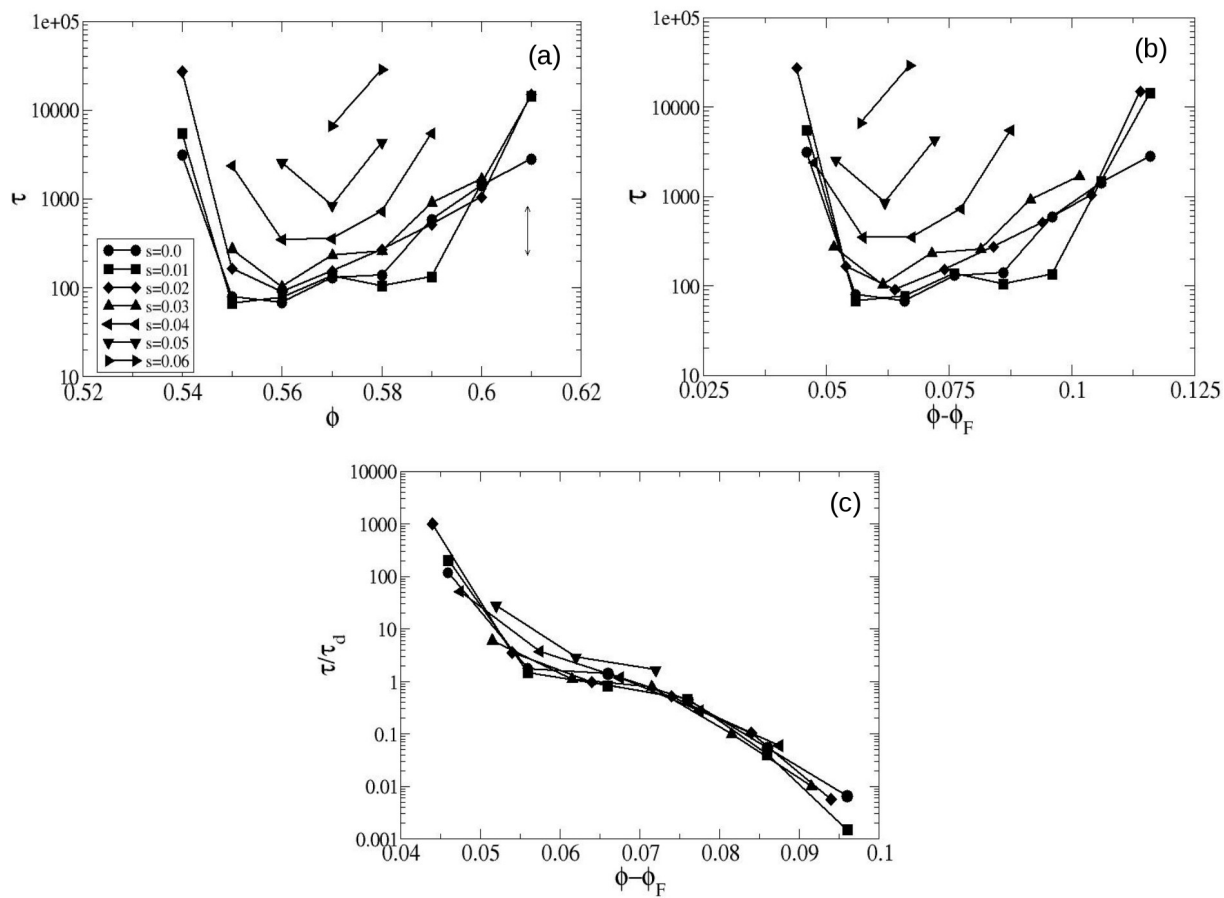

Figure 5

(a) Crystal nucleation times $\tau$ as a function of the packing fraction $\phi$ at the polydispersities $s$ indicated. The arrow indicates the typical uncertainty on a data point. (b) The same data as (a) now plotted versus the supersaturation, $\phi - \phi_F$ where $\phi_F$ is the freezing concentration taken from Fig. 2. (c) The same data again except that the nucleation time $\tau$ is scaled by the time $\tau_d$ that a particle takes to diffuse one diameter.